\documentclass[twocolumn,aps,prl,superscriptaddress]{revtex4}%
\usepackage{color}
\usepackage{amsmath}
\usepackage{amsfonts}
\usepackage{amssymb}
\usepackage{graphicx}%
\providecommand{\U}[1]{\protect\rule{.1in}{.1in}}

\begin{document}
\title{Comment on ``Quantum Szilard Engine"}
\author{Martin Plesch}
\affiliation{Faculty of Informatics, Masaryk University, Brno, Czech Republic}
\affiliation{Institute of Physics, Slovak Academy of Sciences, Bratislava, Slovakia}
\author{Oscar Dahlsten}
\affiliation{Department of Physics, University of Oxford, Clarendon Laboratory, Oxford, OX1
3PU, UK}
\affiliation{Center for Quantum Technology, National University of Singapore, Singapore}
\author{John Goold}
\affiliation{Department of Physics, University of Oxford,
Clarendon Laboratory, Oxford, OX1 3PU, UK}
\author{Vlatko Vedral}
\affiliation{Department of Physics, University of Oxford, Clarendon Laboratory, Oxford, OX1
3PU, UK}
\affiliation{Center for Quantum Technology, National University of Singapore, Singapore}

\maketitle



In a recently published letter \cite{KSLU} the influence of
particle statistics on extractable work in the Szilard engine
\cite{Szilard} was discussed. We point out that the expressions
given there suggest no work extraction is possible in the low
temperature limit if more than two particles are used and thus are
not optimal. We argue that the optimal extractable work is in
general higher and in particular non-decreasing in the number of
particles.

A Szilard engine (SZE) \cite{Szilard} is a hypothetical device
which consists of a cylinder containing a single gas particle,
connected to a heat reservoir at temperature $T$. The engine works
by dividing the cylinder into two parts by a barrier, measuring
the position of the gas particle and extracting work by exploiting
the pressure created by the particle on the barrier.

This process has four stages - insertion of the barrier,
measurement, movement and removal of the barrier. In the classical
case work is extracted only during the movement phase, whereas in
the quantum case work is invested during the insertion process and
is extracted both during the movement and removal phases.

In~\cite{KSLU} the work yield during the removal of the barrier is
calculated in a very specific way. The authors suggest to firstly
lower the potential associated with the barrier to the point where
tunnelling is practically unrestricted. They view any energy that
could potentially be extracted here as lost.  The work is
extracted only in the second phase where the barrier is completely
removed.

We note that with this approach, a problem appears if the barrier
is removed from a position where particles can tunnel from higher
to lower energy levels during the first phase. This happens if the
equilibrium position in terms of the horizontal pressure is not
the middle nor the edge of the box (this requires at least three
particles). Then the ground states on the left and right-hand
sides differ in energy by some non-zero $\Delta E$. In the
tunnelling phase this potentially extractable energy is lost as
the particles
jump to the lower-energy ground state.


The optimal work that can be extracted
has to tend to $0$ as
$kT$ tends to $0$. In the low temperature limit, $kT\ll \Delta E$,
the lost energy is then greater than the optimal extractable work.
It follows that the above protocol will in the low temperature
limit cost rather than yield work.




To illustrate this problem consider a system consisting of three
bosons as depicted in Fig.2 of \cite{KSLU}. After dividing the
piston with a barrier in the middle, with $50\%$ probability we find
3:0 or 0:3, i.e.\ all particles on one side. In this case the optimal procedure
is to move the barrier to the edge of the piston, which can be calculated to yield
$W=2kT\ln(2)$. With $50\%$ probability, however, we have 2:1 or 1:2. In this case the barrier
should be removed from a position where $\Delta E >0$, and this potential work
would be lost using the protocol of \cite{KSLU}.
Altogether the total extractable work would be upper bounded by $W=2kT\ln(2)-\Delta E/2$
and thus negative in the low temperature limit.

To address this issue we correct the expressions used to calculate
the work gained during this process by including the work that can
be gained during the first stage of the removal of the barrier. We
stick to the standard definition of the generalized force as
$F=\frac{\partial E}{\partial\lambda}$ with $\lambda$ being a
parameter of the barrier (e.g. its position during the movement
phase or height during the removal phase) and accept that any such
force can be utilized to perform work. With this assumption (which
is arguably the key point of departure from \cite{KSLU}), the
optimal work is given by $W=kT\ln\frac{Z_f}{Z_i}$ with $Z_f$ and
$Z_i$ the final and initial partition functions respectively.


Following this through, the extractable work in the example mentioned above
becomes $W=2kT\ln(2)$ independently of the measurement
outcome. Results for an arbitrary number $N$ of particles with
different statistics (bosons, fermions and distinguishable
particles) are given in \cite{PDGV}. There we show that work can
be extracted only if the barrier is inserted in a way that
allows for a non-trivial measurement. Then, in the case of distinguishable
particles with a measurement specifying the position of each
individual particle $W_{d}=NkT\ln(2)$. For bosons we get
$W_{b}=kT\ln(N+1)$ and for fermions $W_{f}=kT\ln(2).$ We see that
in all cases the resulting work is non-negative.

{\it Acknowledgements} -- We acknowledge support from the National
Research Foundation (Singapore) and the Ministry of Education
(Singapore). MP acknowledges the support of GA\v{C}R P202/12/1142
and VEGA 2/0072/12.


\end{document}